\documentclass[referee]{raa}            % referee version: for submission

\usepackage{graphicx,times}             %for PS/EPS graphics inclusion, new
\usepackage{natbib}
\usepackage{amssymb,amsmath}
\usepackage{gensymb}
\usepackage{makecell}
\bibpunct{(}{)}{;}{a}{}{,}

\usepackage[driverfallback=dvips,pagebackref=true]{hyperref}
\hypersetup{colorlinks = true, linkcolor = green, anchorcolor = red, citecolor = blue, filecolor = red, urlcolor = red}
\begin{document}

   \title{Numerical study of a 20W class QCW pulsed sodium guide star laser’s performances at five sites in China
%\footnotetext{$*$ Supported by the National Natural Science Foundation of China.}
}
%   \subtitle{I. Place Your Subtitle Here}

   \volnopage{Vol.0 (20xx) No.0, 000--000}      %%preserved for Editor. DOn't remove!
   \setcounter{page}{1}          %%starting page, preserved for Editor. DOn't remove!

   \author{Hong-Yang Li
      \inst{1,3}
   \and Lu Feng
      \inst{1}
   \and Jun-Wei Zuo
      \inst{2}
   \and Qi Bian
      \inst{2}
   \and Bo-Tian Sun
      \inst{4}
   \and Sui-Jian Xue
      \inst{1}
   \and Zhi-Xia Shen
      \inst{1}
   \and Yang-Peng Li
      \inst{1}
   \and Yong Bo
      \inst{2}
   
   }

   \institute{CAS Key Laboratory of Optical Astronomy, National Astronomical Observatories, Chinese Academy of Sciences,
             Beijing 100012, China; {\it Jacobfeng@bao.ac.cn;} {\it lihongyang@nao.cas.cn}\\
        \and
             Technical Institute of Physics and Chemistry, Chinese Academy of Sciences, Beijing 100190, China\\
        \and
             University of Chinese Academy of Sciences, Beijing 100049, China\\
        \and 
             Department of Physics, University of California, Santa Barbara, CA 93106, USA \\
\vs\no
   {\small Received~~20xx month day; accepted~~20xx~~month day}}

\abstract{In the past few years, Chinese astronomical community is actively testing astronomical  sites  for  several  new  optical/infrared  ground-based  telescopes. These site testing campaigns conducted were mainly focused on fundamental performances of the site, such as cloud coverage, seeing, temperature, etc. With  increasing  interests in sodium laser guide star adaptive optics for these new telescopes in the  Chinese astronomical community, it is interesting to investigate the performance of the laser guide star at these sites, especially considering that the sodium laser guide star’s on-sky performance is significantly influenced by sites’ local performances, such as geomagnetic field, sodium layer dynamics, density of air molecule, etc. In this paper, we studied sodium laser guide star’s performance of a 20W class Quasi-CW pulsed laser developed by TIPC with numerical simulation for five selected sites in China.
}

   \authorrunning{H.-Y. Li, L. Feng \& J.-W. Zuo}            %author_head in even pages
   \titlerunning{Simulation of 20W class Quasi-CW pulsed laser}  % title_head in odd pages

   \maketitle
%% The author head (on even pages) and the title head (on odd pages) will be
%% automatically extracted from \author{} and \title{}. Whenever the title is too long,
%% you will be asked to supply a shorter one by inserting either \authorrunning{} or
%% \titlerunning{} before \maketitle. Anyway, you can specify your own heads.
%%
%%
%% Note: In the following text body of your manuscript, please note several differences from
%%       other major journals:
%% (1) \subsection{Please Capitalize the First Letter of Each Notional Word in Subsection Title}
%% (2) Please Capitalize the First Letter of Each Notional Word in all tables' captions

%
%________________________________________________ sections below
%
\section{Introduction}           %% first-level sections will be auto-capitalized
\label{section: introduction}

As a subsystem of Adaptive Optics (AO) system, sodium laser guide star system can provide a stable and reliable reference source for telescopes' AO systems to correct atmospheric turbulence (\citealt{Hardy+1998}). Sodium laser guide star systems have been utilized in most of those highly-productive large ground-based telescopes , such as the Keck Telescope (\citealt{Chin+etal+2016}), the European Southern Observatory’s (ESO) Very Large Telescope (\citealt{Calia+etal+2014}), the twin Gemini Telescopes, the Gran Telescopio CANARIAS (GTC) \textit{et cetera}, and such system is also designed in the first place for those next generation 30m class ground-based telescopes, such as the Thirty Meter Telescope, the European Extremely Large Telescope (E-ELT) and the Giant Magellan Telescope (GMT). 

One of the most important future astronomical projects within the Chinese astronomical community, the 12m diameter Large Optical/infrared Telescope (LOT) is also planned to have AO system installed as soon as the telescope is constructed (\citealt{Feng+etal+2020}). By far, several excellent candidate sites for this large telescope and several smaller telescopes to be built in near future have been investigated in the western part of China, such as Tibet, Xinjiang, Qinghai, Sichuan, etc (\citealt{Feng+Hao+2020}). Site monitoring scientists have made great efforts in evaluating  fundamental astronomical performances of these sites, including cloud coverage, seeing, meteorological parameters, preciptical water vapor, \textit{et cetera}  (\citealt{Liu+etal+2012,Liu+etal+2015,Liu+etal+2016,Yao+etal+2013,Qian+etal+2015,Wang+etal+2015,Wu+etal+2016}), but performances relating to sodium laser guide star's on-sky performances at these candidate sites, like the geomagnetic field, sodium layer dynamics and so on, are not being monitored due to the difficulties and costs for such measurements. Therefore, we investigated the performance of sodium guide star at certain typical sites in China in advance using our model combining with data from remote sensing satellites as well as other atmospheric models, to quantify the influence of some parameters related to sites on the performance of sodium guide star, such as geomagnetic field and sodium layer dynamics. The results of this study may shed lights on the future application of sodium guide stars in the western part of China.

\begin{figure}
\centering
\includegraphics[width=\linewidth]{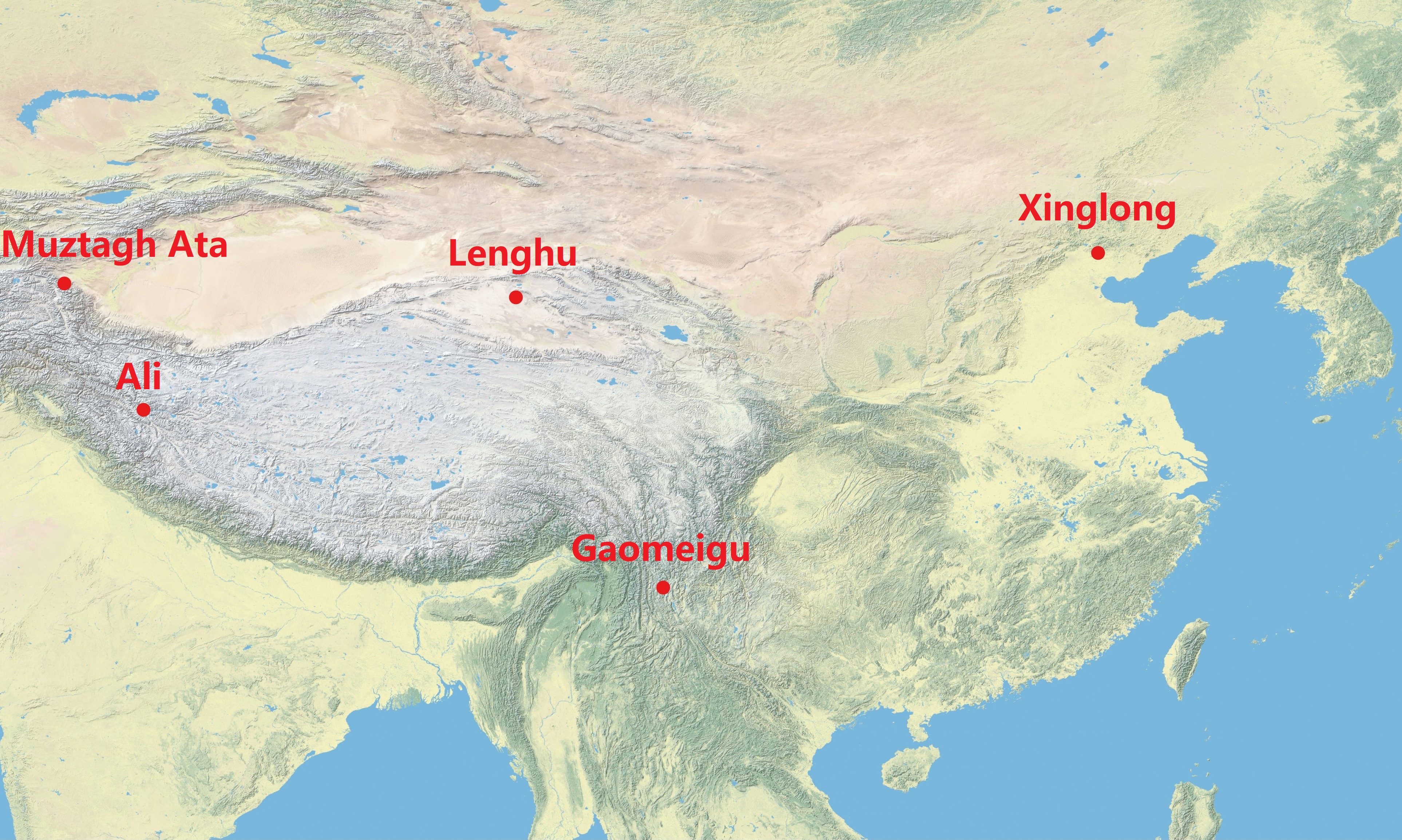}
\caption{The location of the five sites in China (Xinglong, Lenghu, Muztagh Ata, Ali and Gaomeigu)}
\label{figure: Location of five sites}
\end{figure}

For comparative simulation study for the on-sky performance of sodium LGS, we selected five sites for this study: Ali, Lenghu, Muztagh Ata, Gaomeigu in western China and Xinglong (an important site in eastern China) as a comparison. Figure \ref{figure: Location of five sites} shows the location of the five sites. The Ali site is located in the southwest of Tibet autonomous region with an altitude of 5100m. It is one of the three candidate sites for the 12-meter Large Optical/infrared Telescope (LOT) (\citealt{Yao+etal+2013}). The Muztagh Ata site is also one candidate site for LOT, which is located in the Pamir Plateau, in the southwest of Xinjiang Province of China, and more information about the site can be seen from \cite{Xu+etal+2020}. The Lenghu site is the site for the future 2.5m Wide Field sky Survey Telescope(WFST) and it is located in the northwest border of Qaidam Basin with an average altitude of 2800m. The Gaomeigu site is located in Lijiang, Yunnan province with an altitude of 3200m, which has the largest general-purpose optical telescope, Lijiang 2.4-meter Telescope (LJT), in China. Xinglong site, located at the highest latitude among the selected sites, has a history of more than 50 years and possesses two important astronomical observation instruments, the LAMOST and the 2.16m Telescope.

The laser that we chose to simulate in this study is a 20W Quasi Continuous Wave (QCW) pulsed laser from the Technical Institute of Physics and Chemistry (TIPC) with whom we have been cooperating both in experiments and simulation study of sodium laser guide star for years. Significant amount of lab/field tests data have been gathered  (\citealt{Wei+etal+2012,Jin+etal+2014,Otarola+etal+2016,Bian+etal+2016a,Bian+etal+2016b,Bian+etal+2017,Bian+etal+2020}), and were used to verify our model previously (\citealt{Jin+etal+2015, Feng+etal+2016}). Although we are confident that our model would be suitable to predict LGS generation performances for other types of sodium laser, such as Continuous Wave (CW) laser, Macro-micro pulsed laser (\citealt{Li+etal+2021}), we would like to present prediction results for such lasers at different sites until sufficient field tests results have been gathered to verify our model.

In the rest of the paper, we will describe parameters related to the laser and sites in Section \ref{section: method}, and the simulation results will be presented in Section \ref{section: results}.  

%% Authors can give a citation as 'Michel et al. 1992'.
%% You may also use \cite, \citep and \citet for citation, and use Table~1 or Figure~1
%% and so forth. Using \ref and \label for cross-references of Tables/Figures
%% is a good way in adjusting/adding/removing text, tables or figures.

\section{METHOD}
\label{section: method}

In this paper, we used a simulation tool called Photon Return Simulation program (PRS) (\citealt{Feng+etal+2015,Jin+etal+2015,Li+etal+2021}) to predict the on-sky performance of 20W Quasi-CW pulse laser at aforementioned sites in China. The simulation is a Monte Carlo simulation based on rate equation. During the simulated time period (for example 1s), the state of each atom under the influence of laser light field as well as other different physics such as geomagnetic field influence, light pressure, D2b repumping are being tracked for every 30 nanosecond. The sampling time is short enough to simulate a 100$\mu$s QCW laser pulse. We simulate this one second process repeatedly for about 100 thousand times, which is sufficient for the result of the Monte Carlo simulation to converge and to calculate the mean value of coupling efficiency as in the Lidar equation as a representative for each laser power density. After that, we could obtain the coupling efficiency versus the projected on-sky power density curve. In this section, we will introduce the sources and values of these parameters.

\subsection{The parameters of TIPC laser}
\label{section: sub: laser parameters}
\begin{figure}
\begin{minipage}[t]{0.5\linewidth}
\centering
\includegraphics[width=\linewidth]{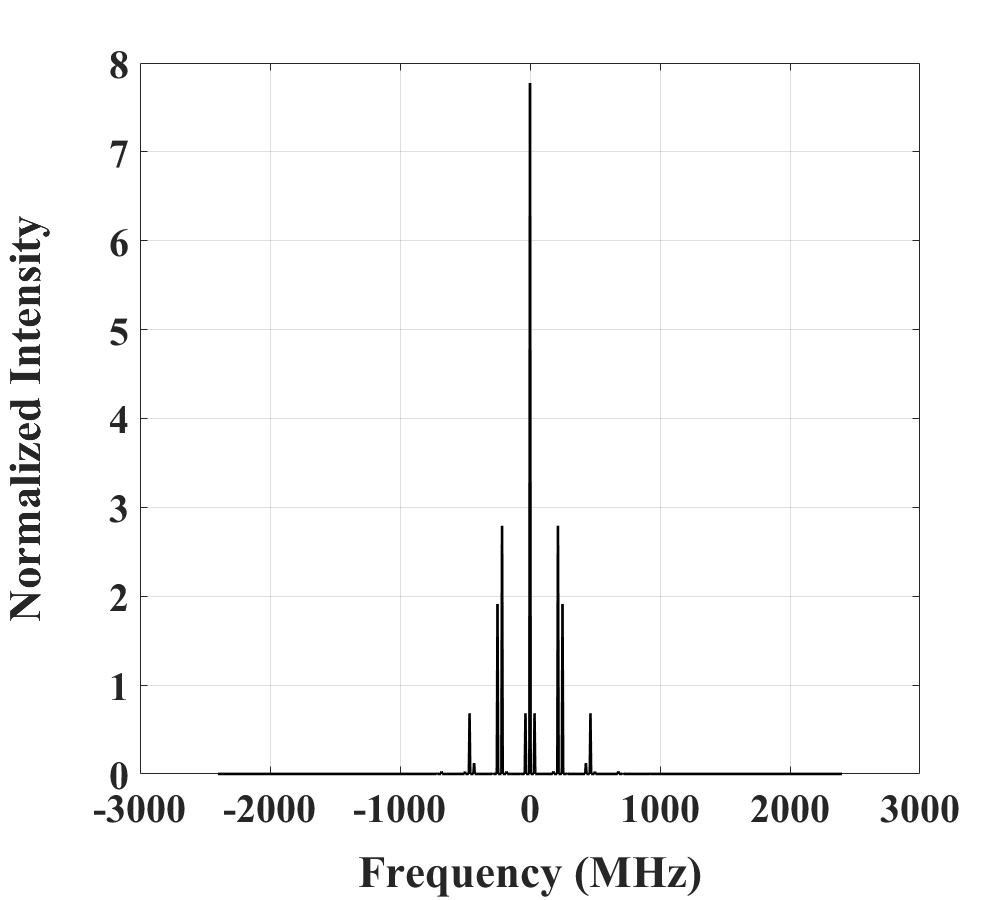}
\end{minipage}%
\begin{minipage}[t]{0.5\linewidth}
\centering
\includegraphics[width=\linewidth]{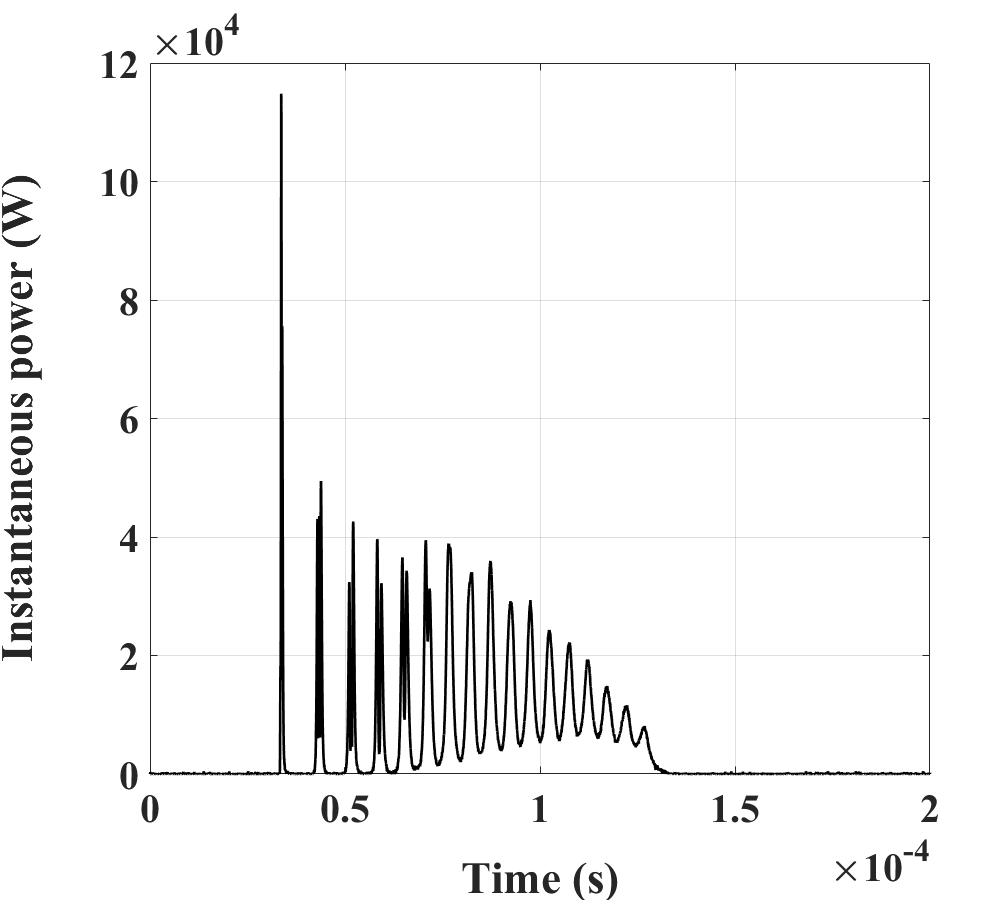}
\end{minipage}
\caption{Spectrum format and single pulse format of TIPC 20W Quasi-CW pulse laser. (left) The width (FWHM) of spectrum is 300MHz and the width (FWHM) for each longitudinal modes is 5MHz.  (right) The 100 $\mu s$ single pulse format of the laser}
\label{figure: spectrum and temporal format}
\end{figure}

The spectrum and pulse format of sodium laser plays an important role in predicting the return flux of sodium guide star (\citealt{Kibblewhite+2000}). As described previously, the laser investigated in our study is a 20W quasi-CW pulse laser from TIPC. Main parameters of the laser are summarized in Table \ref{table: laser parameters list}. The spectrum and single pulse format of the laser are shown in Figure \ref{figure: spectrum and temporal format}. The central wavelength of the laser is at sodium D2a line, 589.159nm, which corresponds to 0 MHz in Figure \ref{figure: spectrum and temporal format}. The Full Width Half Maximum (FWHM) of the spectrum is about 0.3 GHz. The spectrum of the laser contains several longitudinal modes and the distances between adjacent longitudinal modes are equal which is about 225 MHz. 

\begin{table}[]
\centering
\begin{tabular}{|l|l|}
\hline
\textbf{Parameter name}                                                                             & \textbf{Value}             \\ \hline
Center wavelength                                                                          & $589.159$ nm        \\ \hline
Linewidth                                                                                  & $\sim0.3$ GHz     \\ \hline
Longitudinal mode interval                                                                 & $\sim225$ MHz           \\ \hline
Longitudinal mode width (FWHM)                                                                & $5$ MHz           \\ \hline
Pulse repetition rate                                                                      & $600$ Hz            \\ \hline
Pulse width                                                                                & $\sim100$ $\mu$s \\ \hline
Average power                                                                              & $20$ W              \\ \hline
%Beam quality                                                                               & $1.4$             %  \\ \hline
%Divergence angle                                                                           & $1$ mrad          %  \\ \hline
Polarization                                                                               & circular          \\ \hline
D2b repumping factor                                                                       & $ 15\%$ 
\\ \hline
%\begin{tabular}[c]{@{}l@{}}Laser guide star on sky spot size\\ (FWHM at 92km)\end{tabular} & 1$\sim$1.5 arcsec \\ \hline
\end{tabular}
\caption{Laser parameters for TIPC 20 Watt Quasi-CW pulsed sodium laser}
\label{table: laser parameters list}
\end{table}

\subsection{Sodium atom column density}
\begin{figure}
\begin{minipage}[t]{0.5\linewidth}
\centering
\includegraphics[width=\linewidth]{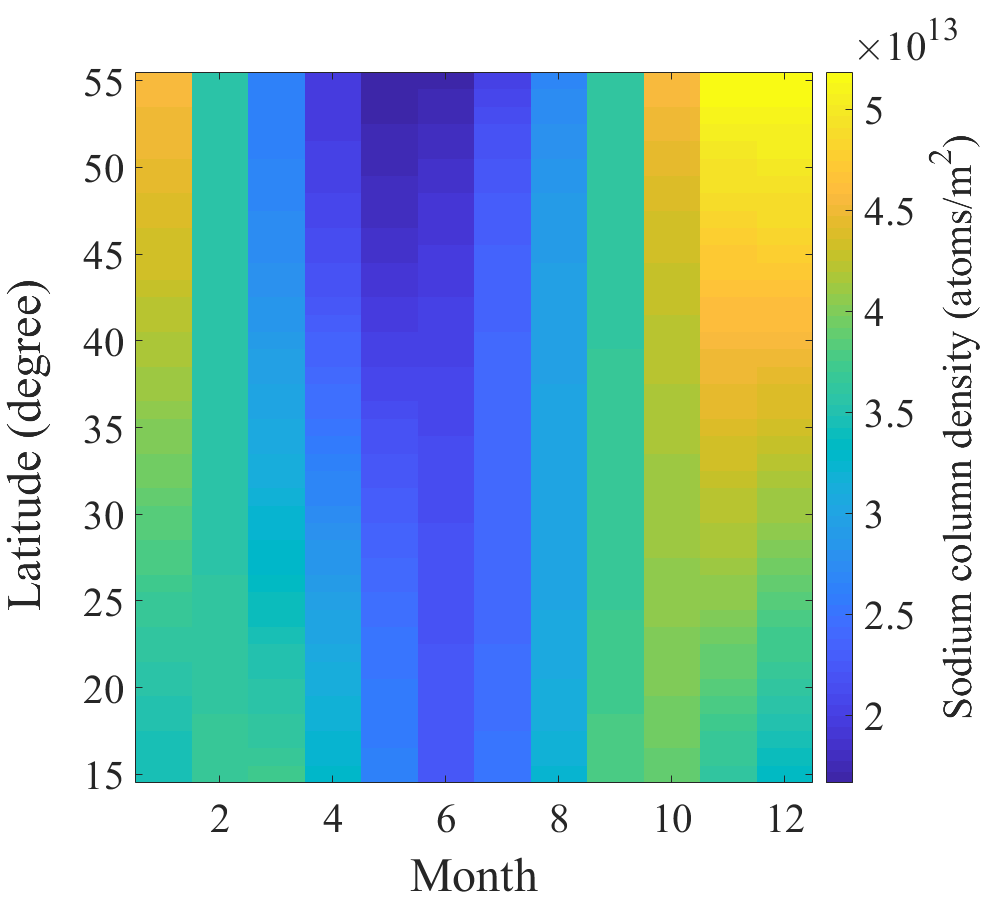}
\end{minipage}%
\begin{minipage}[t]{0.5\linewidth}
\centering
\includegraphics[width=\linewidth]{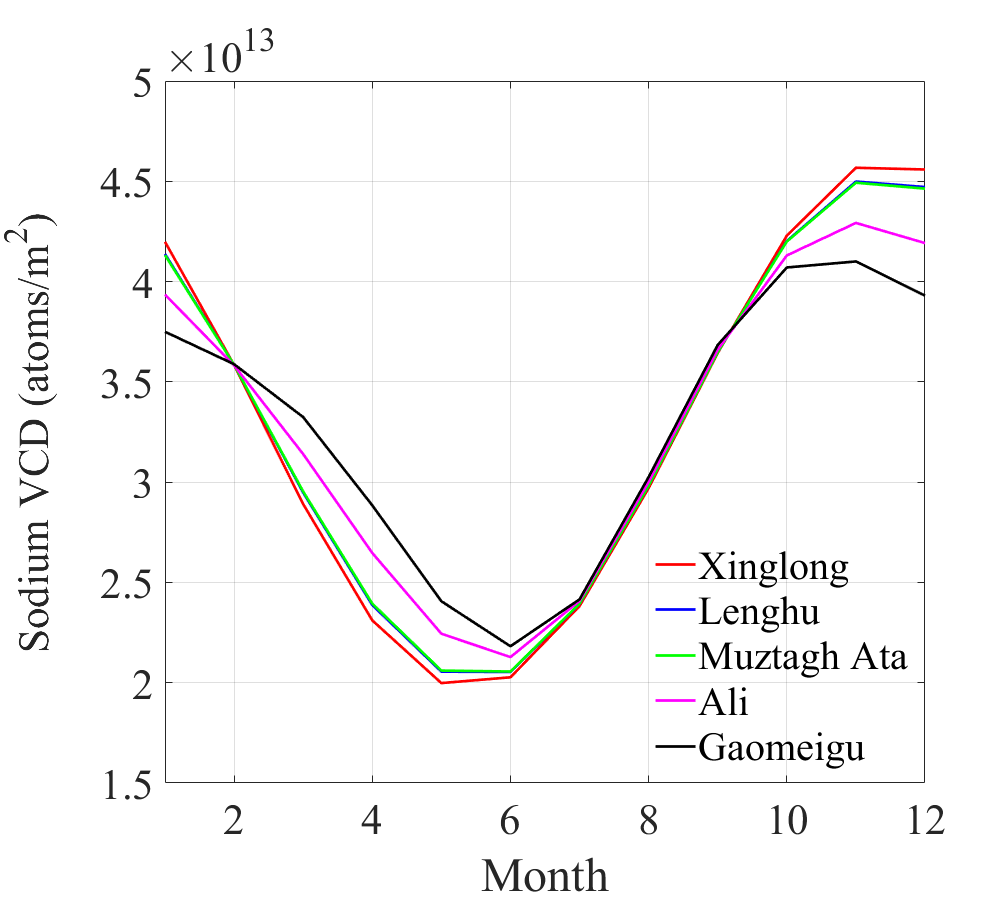}  
\end{minipage}
\caption{(left) Sodium column density changes with month and latitude (15 \degree N-55 \degree N) in China and is basically higher in winter and lower in summer. (right) Five typical sites' sodium column density varies with month. The densities at Lenghu and Muztagh Ata are almost the same, so the curves almost coincide.}
\label{figure: sodium column density}
\end{figure}

Photon return number is closely related to the sodium atom column density in mesosphere, while the sodium Vertical Column Density(VCD) will change obviously with latitude (\citealt{Langowski+etal+2017}). We utilize the empirical formula of sodium column density proposed by Fussen, which is based on remote sensing data from GOMOS (\citealt{Fussen+etal+2004,Fussen+etal+2010}). The satellite is designed for measuring column densities of molecules or atoms including sodium atom in the atmosphere. Its results are further explored to show the variation of the sodium VCD with different latitudes and different months. Based on these data, Fussen formed an empirical formula of sodium column density against location and month as shown in Equation \ref{equ: column density}. In this equation, N represents sodium density for a certain month m (January is m=0) and latitude $\phi$ (in radians for Eq. \ref{equ: column density}). Other parameters in Equation \ref{equ: column density} are listed in Table \ref{table: column density equation parameters}. According to this formula, we calculated the sodium column density in the mesosphere in different months in China for the latitude ranging from 15\degree N to 55\degree N , and the detailed results are visualized from Figure \ref{figure: sodium column density} (left). In general, the sodium column density is higher in winter and lower in summer at the same latitude. With the increase of latitude, it increases in winter and decrease in summer, so the difference between the two seasons raises with latitude. 

The sodium VCD reaches the highest in November, the lowest in May or June at the selected sites in China, as shown in Figure \ref{figure: sodium column density} (right). The differences of sodium VCD between winter and summer are relatively large. For example, even for Gaomeigu site-the site with smallest annual difference among the sites-the sodium VCD is 87.9\% higher in November than in June: in November, it can reach $4.1\times10^{13}$ (atoms/$m^{2}$), whereas in June, it is only $2.2\times10^{13}$ (atoms/$m^2$). Moreover, the difference of the sodium VCD between different sites is relatively large from November to January next year, while the difference between June and October is very small.

\begin{gather}
    N(m,\phi)[cm^{-2}] = t_0 +t_1 \cos(\frac{2\pi}{12}m + t_2) + t_3(\phi+\frac{\pi}{2})(\phi-\frac{\pi}{2})\cos(\frac{2\pi}{6}m+t_4) \notag\\
    t_{i\leq1} = f_i(a_0+a_1\phi+a_2\phi^2+a_3\phi^3) \notag\\
    t_0 = 3.28\times10^9
\label{equ: column density}
\end{gather}

\begin{table}[]
\centering
\begin{tabular}{|l|l|l|l|l|l|}
\hline
i & $f_i$                                        & $a_0$    & $a_1$    & $a_2$    & $a_3$    \\ \hline
1 & $1\times10^9$ & 0.1282  & 1.549   & 0.1780  & 0.03511 \\ \hline
2 & 1                                           & 0.4017  & 0.8216  & -0.1282 & -0.2980 \\ \hline
3 & $1\times10^9$ & -0.2630 & 0.1121  & 0.6355  & -0.3566 \\ \hline
4 & 1                                           & -1.5635 & -3.0526 & 1.3802  & 1.7637  \\ \hline
\end{tabular}
\caption{Parameters for Equation \ref{equ: column density}.}
\label{table: column density equation parameters}
\end{table}

\subsection{The effect of geomagnetic field}
The Larmor precession caused by the geomagnetic field will result in the M state of sodium atoms to change periodically with time (\citealt{Moussaoui+etal+2008,Moussaoui+etal+2009}). The period depends on the strength of the geomagnetic field and the angle between the pointing direction of the laser and the geomagnetic field vector. Since the transition cross section of the sodium atom is different when it is at different M ground state, the periodical shift due to Larmor precession will cause a degradation of the overall coupling efficiency for LGS generation. The strength and direction of the geomagnetic field vector will change with the location of the site. The coupling efficiency and photon flux will change accordingly. This will lead to some differences in the coupling efficiency and photon return number at these sites, so it is necessary to determine the strength and direction of the geomagnetic field at these sites. In Table \ref{table: sites parameters}, we listed the sites' local geomagnetic field information, including declination/inclination angle and the strength of the geomagnetic field vertor, which is from the World Magnetic Model (WMM). 

The pointing direction of laser will change as the telescope tracks the observation target. Therefore, the angle between the laser and the geomagnetic field will also change accordingly. In our study, we evaluate the on-sky performance variation due to geomagnetic field effects. When we were analyzing the on-sky performance with different projected laser power for different months, we only simulate in the zenith direction because of two reasons: 1. to easily compare with LGS field tests, since for these field tests, the projected power is deliberately varied to check against model prediction, and it is normally only conducted in the zenith without any requirement for complex laser transfer optics for pointing mechanism, 2. to simplify the calculation process. Only when discussing the annual mean on-sky performance in different directions, we did the simulation for all five sites with the full power of the laser at 20W for all elevation/azimuth angle combinations.  

%While, we only simulate the zenith direction \textcolor{red}{[of five sites]} in this paper, for two reasons. One is to simplify the calculation and facilitate the comparison of simulated results between different sites. It’s necessary to choose a representative direction for simulation and it’s obvious that the zenith \textcolor{red}{[at different sites]} is optimum. The other is to compare with the results of field tests in the future, because the zenith direction is the preferred direction in field tests.

\subsection{Temperature and atmospheric molecular density}
Atmosphere molecular density and temperature in the mesosphere will influence the collision possibility between sodium atoms and atmospheric molecules/atoms and then affect the coupling efficiency and the photon return number (\citealt{Kibblewhite+2000,Kibblewhite+2008}). The values of the two parameters are related with the location of a site and can be obtained by querying MSISE-90 model from CCMC.

\begin{table}[]
\centering
\begin{tabular}{|c|c|c|c|c|c|c|}
\hline
Site  & Latitude  & Longitude  & Altitude  &  \makecell[c]{Declination \\ (+E $|$ -W)}  &  \makecell[c]{Inclination \\ (+D $|$ -U)}  &  \makecell[c]{Strength of Local \\ geomagnetic field (nT)}    \\ \hline
Xinglong & 40.39\degree N & 117.58\degree E  & 960m  & -7.05\degree  &  59.33\degree  & $5.20\times10^{4}$    \\ \hline
Lenghu & 38.45\degree N  & 93.20\degree E  & 2800m    & 3.58\degree  &  57.99\degree  & $5.22\times10^{4}$    \\ \hline
Muztagh Ata & 38.28\degree N  &  74.90\degree E  & 4526m   & 3.56\degree  &  57.99\degree & $5.03\times10^{4}$    \\ \hline
Ali & 32.31\degree N   & 80.05\degree E  & 5100m   & 1.61\degree  &  50.33\degree  & $4.80\times10^{4}$    \\ \hline
Gaomeigu & 26.71\degree N & 100.03\degree E  &  3200m   & -1.11\degree  &  41.59\degree  & $4.61\times10^{4}$    \\ \hline
\end{tabular}
\caption{Parameters related to sites}
\label{table: sites parameters}
\end{table}

\section{RESULTS}
\label{section: results}

\subsection{Coupling efficiency variation under circular polarization at the zenith}
\label{section: sub: coupling efficiency at five sites}
The coupling efficiency of circularly polarized light is higher than that of linearly polarized light due to the larger scattering cross section of the circular polarization. For this reason, all LGS lasers are circularly polarized. In the simulation we take circular polarization as a default to simulate the performance of the TIPC laser. 

\begin{figure}
\centering
\includegraphics[width=\linewidth]{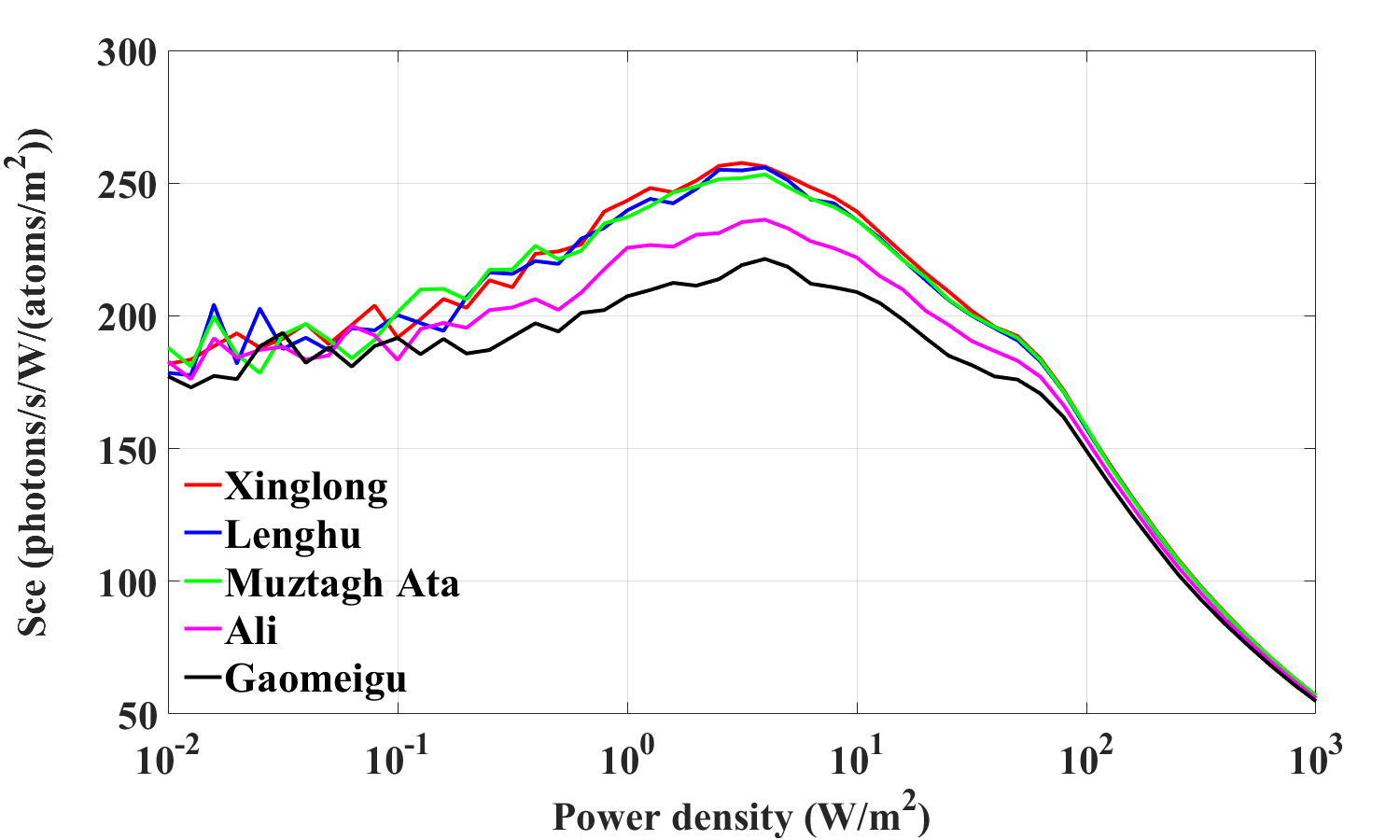}
\caption{Coupling efficiency against projected power density on sky (92km) in zenith direction}
\label{figure: coupling efficiency at five sites}
\end{figure}

The comparison of the coupling efficiencies at the zenith with different power densities is shown in Figure \ref{figure: coupling efficiency at five sites}. The range of power density on sky (X-axis) is from 0.01 to 1000 $W/m^2$ and the Y-axis is the coupling efficiency. The curves of different colors represent the variations of coupling efficiency against projected power densities at different sites. Coupling efficiencies at Xinglong, Lenghu and Muztagh Ata are almost identical, which are higher than Ali and Gaomeigu. This is mainly due to the influence of geomagnetic field. When the direction of laser and geomagnetic field vector is parallel, lamor precession caused by geomagnetic field effect is the least; when the two is vertical, the influence of lamor precession is the strongest. From Table \ref{table: sites parameters}, we can see that the angles of geomagnetic field and the zenith at the former three sites are almost the same, and the geomagnetic field strength is similar. Both of them are higher than that of the latter, which explains why the coupling efficiency of Ali and Gaomeigu at zenith is lower.

\subsection{Photon return efficiency in zenith}
\label{section: sub: photon return efficiency}

In the zenith direction, the sodium column density in the mesosphere will change with month and latitude of the site, and the coupling efficiency will change with the location of sites and power density. Therefore, we use the product of the sodium column density and the coupling efficiency to investigate  the change of the photon return number with the power density in different months at different sites. We temporarily call the product of the two as the photon return efficiency in zenith.

\begin{figure}
\begin{minipage}[t]{0.5\linewidth}
\centering
\includegraphics[width=\linewidth]{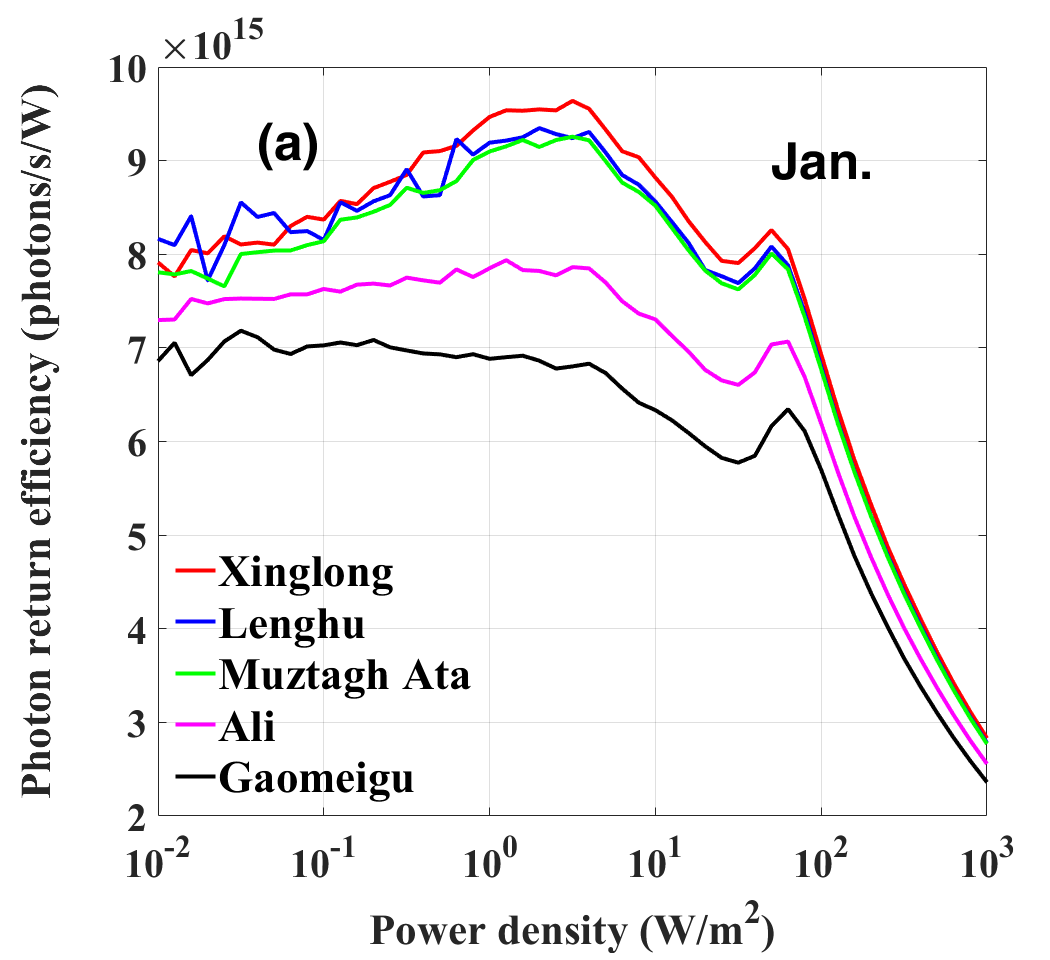}
\end{minipage}%
\begin{minipage}[t]{0.5\linewidth}
\centering
\includegraphics[width=\linewidth]{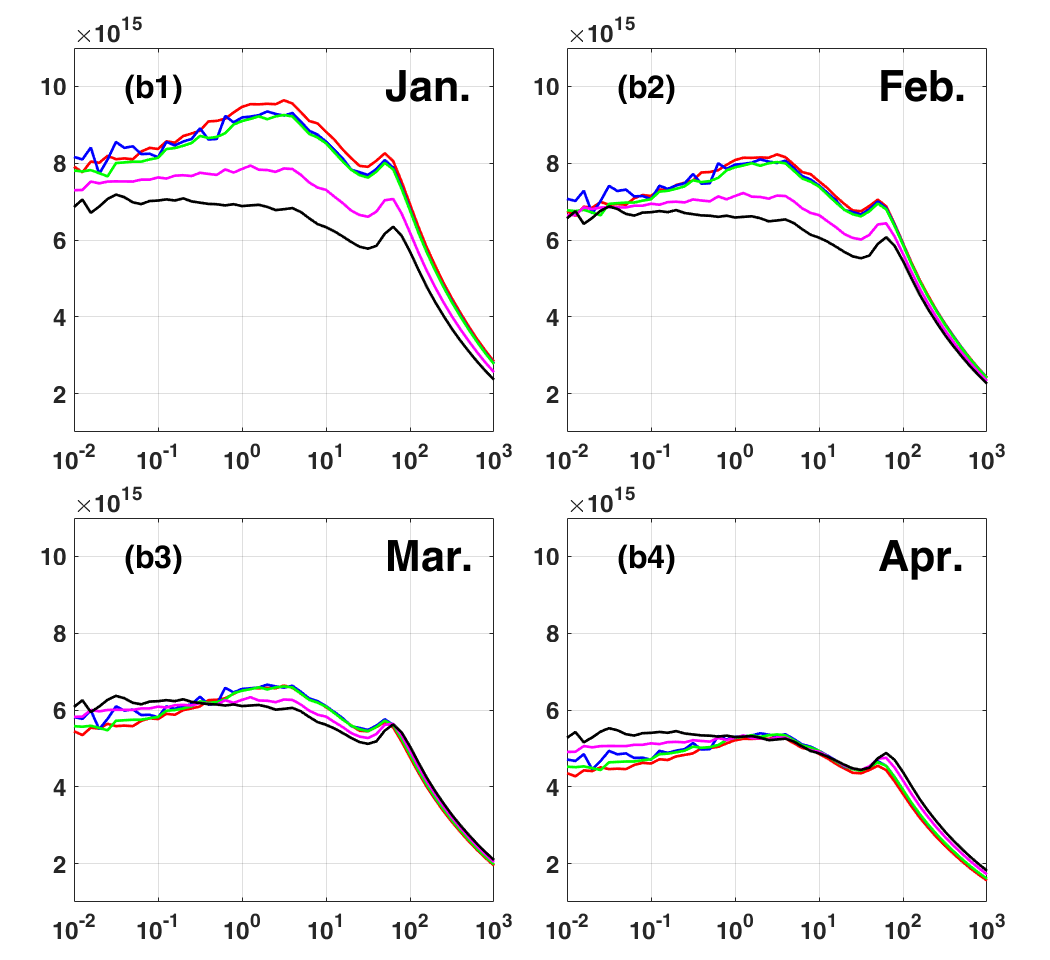}
\end{minipage}
\begin{minipage}[t]{0.5\linewidth}
\centering
\includegraphics[width=\linewidth]{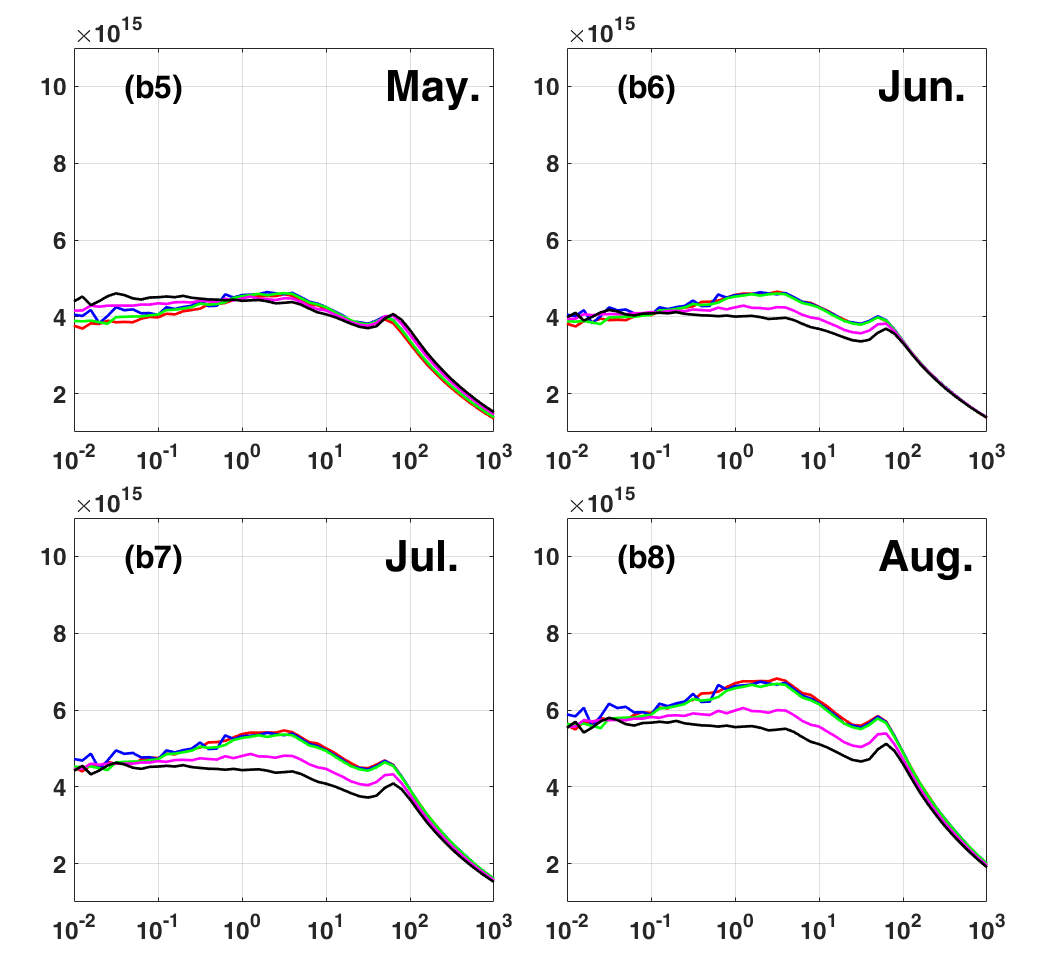}
\end{minipage}%
\begin{minipage}[t]{0.5\linewidth}
\centering
\includegraphics[width=\linewidth]{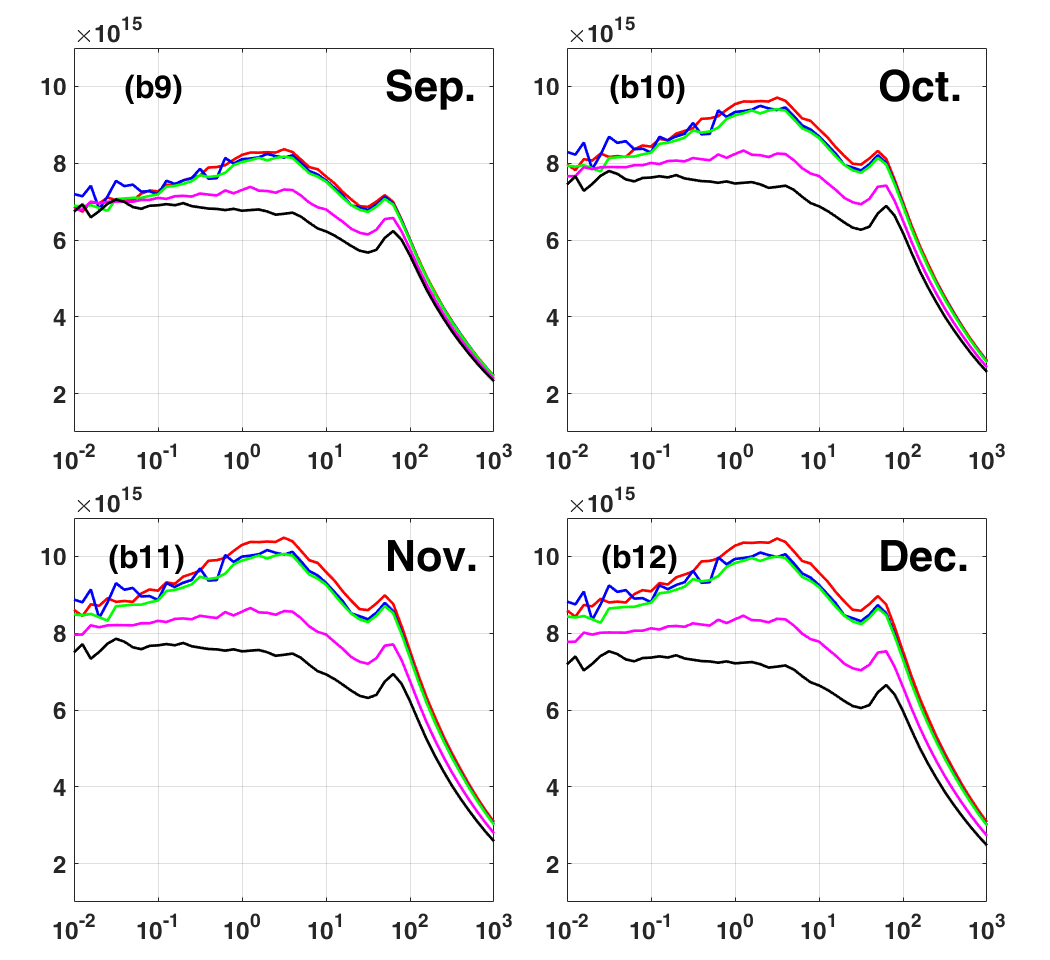}
\end{minipage}
\caption{Photon return efficiency variation against power density on sky (92km) from Jan. to Dec (b1-b12).}
\label{figure: photon return efficiency}
\end{figure}

Such photon return efficiencies from January to December are shown in Figure \ref{figure: photon return efficiency}. Figure \ref{figure: photon return efficiency}a shows the photon return efficiency variation with the power density on sky in January. The curves with different colors represent the efficiencies for different sites, which are the same as in Figure \ref{figure: coupling efficiency at five sites}. Figure \ref{figure: photon return efficiency} b1-b12 shows the change of photon return efficiency from January to December and the ranges of value for Y axis in these figures are the same for easy comparison. We can summarize Figure \ref{figure: photon return efficiency} as follows:

\begin{itemize}
    \item In different months, the photon return efficiency of a site varies greatly. For example, the maximum photon return efficiency of Xinglong in November is 129\% higher than that in June: in June, the photon return efficiency is $4.65\times10^{15}$ (photons/s/W), but in November is $10.65\times10^{15}$ (photons/s/W). This is mainly caused by the large change of sodium column density with the month;
    \item In spring and summer (from Mar to August), the photon return efficiencies of different sites are similar and at a low level. While in autumn and winter (from September to February), there is large differences between the photon return efficiency of five sites. The higher the latitude of site, the higher the photon return efficiency. There are two main reasons for this phenomenon: one is that the coupling efficiency increases with latitude of site due to the influence of geomagnetic field; the other is that the sodium column density keeps at high level in autumn and winter.
\end{itemize}

\subsection{Returned photon flux}
\label{section: sub: photon return number at five sites}

%\textcolor{red}{[Photon return number determines the brightness of the guide star that the laser can produce, which is exactly what astronomers are most concerned about. The relationship between the coupling efficiency and photon return number on sky is shown in Equation 2, where P(x,y) is the distribution of power on sky, $C_{Na}$ is the sodium column density, Sce is the coupling efficiency and T is the atmospheric transparency. The mean power of TIPC laser is 20W. Assuming that the beam diameter at 92km is 1m and 3m respectively, the atmospheric transparency is 75\%, and power on sky is Gaussian distribution, the photon return number on sky at five sites at the zenith in different months can be calculated, as shown in Figure 6. Figure 6 left and right show photon return number of five sites changing with month when the beam diameter on sky is 1m and 3m respectively. The photon return numbers at Xinglong, Lenghu and Muztagh Ata are very close and higher than those of Ali and Gaomeigu, which is more obvious in winter. This is mainly due to the influence of geomagnetic field effect. For the same site, the photon return number on sky with a 3m beam diameter is more than that with a 1m diameter.]}

\begin{figure}
\begin{minipage}[t]{0.5\linewidth}
\centering
\includegraphics[width=\linewidth]{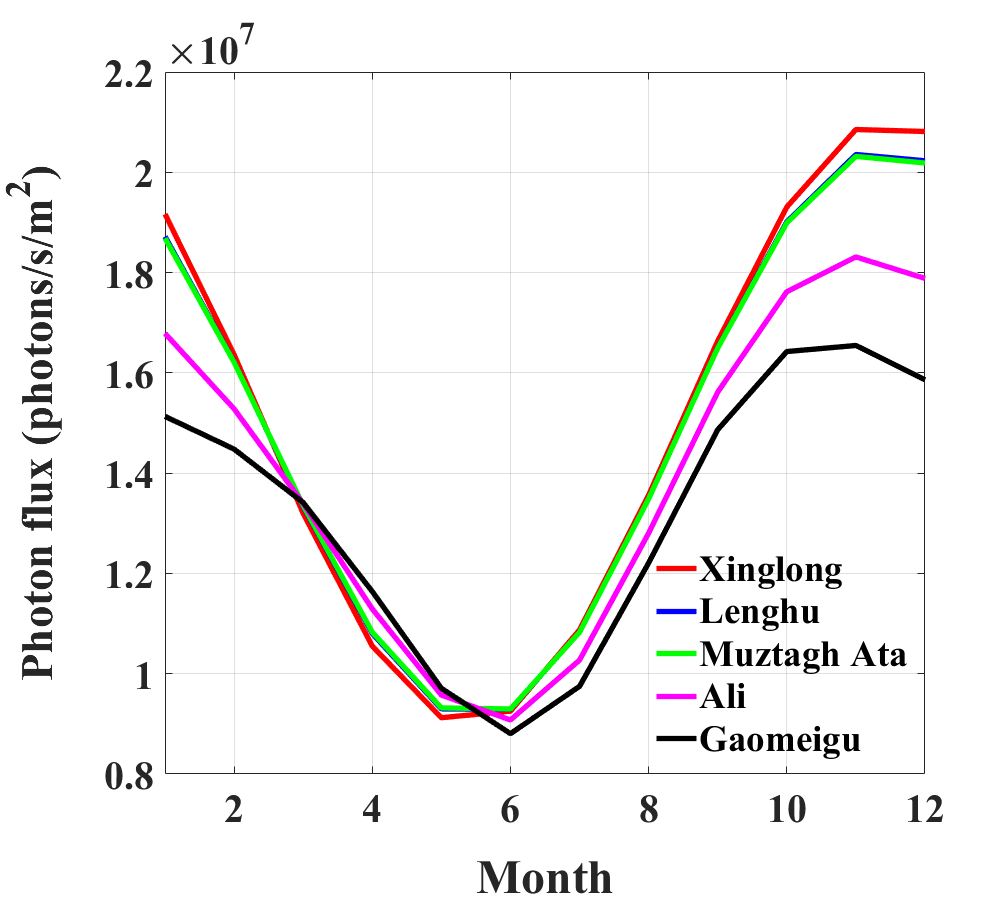}
\end{minipage}
\begin{minipage}[t]{0.5\linewidth}
\centering
\includegraphics[width=\linewidth]{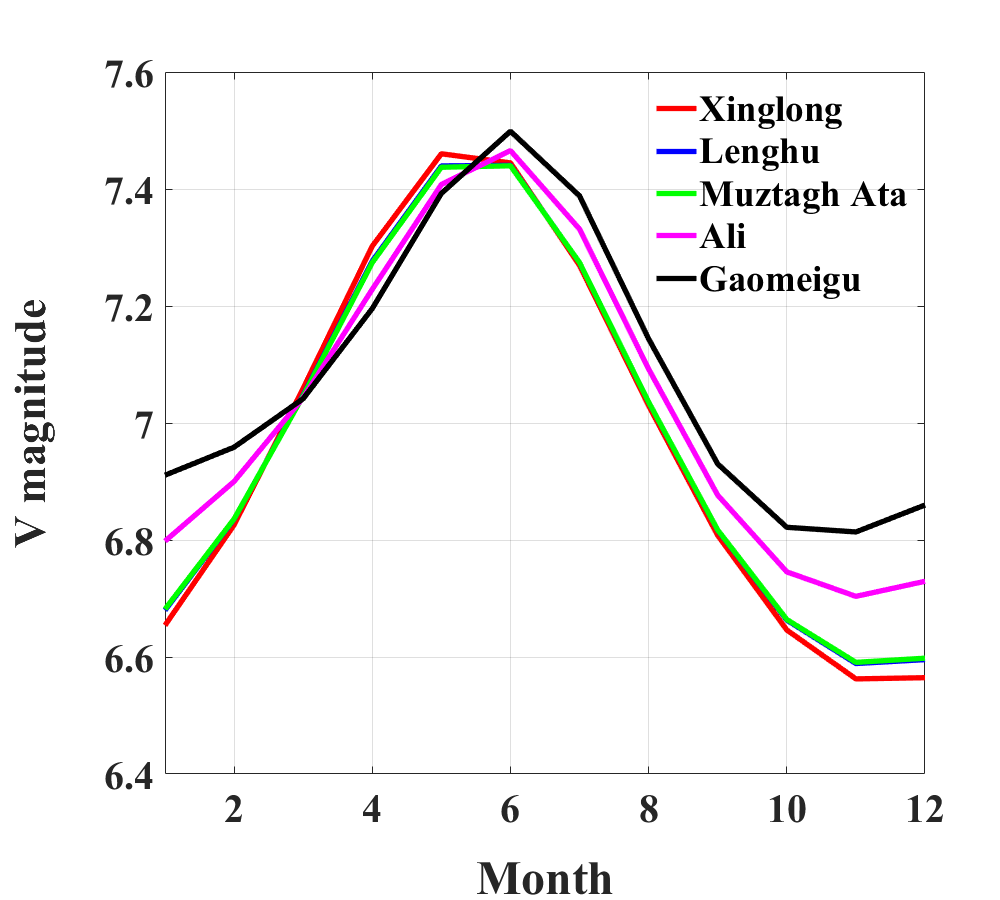}
\end{minipage}
\caption{Photon flux variation with month (left) and corresponding seasonal distribution of V-band magnitude of sodium laser guide star (right)}
\label{figure: photon flux and V magnitude}
\end{figure}

The returned photon flux determines how much photons could reach the telescope, which is exactly what astronomers are most concerned about. The relationship between the coupling efficiency and the photon flux can be described by equation \ref{equ: Lidar equation} from Holzlöhner’s paper (\citealt{Rochester+etal+2010}), where Sce is the coupling efficiency, $\Phi$ is the photon flux received by the telescope (unit photons/s/$m^2$), L is the vertical distance from the primary of the telescope to the sodium layer, P is the laser power, T is the atmospheric transparency, X = 1/($sec(\theta)$) and $\theta$ is the laser launch zenith angle, $C_{Na}$ is the sodium column density. Assuming that the atmospheric transparency at zenith is 90\%, the seeing is 1.5” and beam quality ($M^2$) of TIPC laser is 1.4, the diameter of laser spot size at 92km on sky is about 0.737m(1.652”), and the mean power density at zenith is about 42.2 W/$m^2$. The coupling efficiency can be obtained from the curve of the coupling efficiency against power density(Figure \ref{figure: coupling efficiency at five sites}). According to Equation \ref{equ: Lidar equation}, the photon flux in zenith for different months can be calculated, as shown in Figure \ref{figure: photon flux and V magnitude} (left). In general, the photon flux is higher in winter and lower in summer, and the flux variations at Xinglong, Lenghu and Muztagh Ata are identical and higher than that of Ali and Gaomeigu, which is more pronounced in winter. This is mainly due to the influence of the seasonal distribution of the sodium layer column density. According to the relationship of photon flux and magnitude of an object (Astronomical Magnitude Systems: https://www.cfa.harvard.edu/~dfabricant/huchra/ay145/mags.html), we calculated the corresponding Johnson V-band magnitude variation of the sodium guide star with month, shown in Figure \ref{figure: photon flux and V magnitude} (right).  The highest V magnitude is up to 6.56 in November at Xinglong and the lowest is about 7.50 in June at Gaomeigu.

%\begin{gather}
%    N_{Na} = \iint P(x,y) \cdot C_{Na} \cdot Sce \cdot T \,dx \,dy
%\label{equ: photon return number on sky}
%\end{gather}

\begin{gather}
    \Phi = \frac{Sce \cdot P \cdot (T)^{2X} \cdot C_{Na} \cdot X}{L^2}
\label{equ: Lidar equation}
\end{gather}

\subsection{Coupling efficiency variation with direction of the laser}
\label{section: sub: Coupling efficiency variation with direction of the laser}

\begin{figure}
\begin{minipage}[t]{0.5\linewidth}
\centering 
\text{Xinglong}
\includegraphics[width=\linewidth]{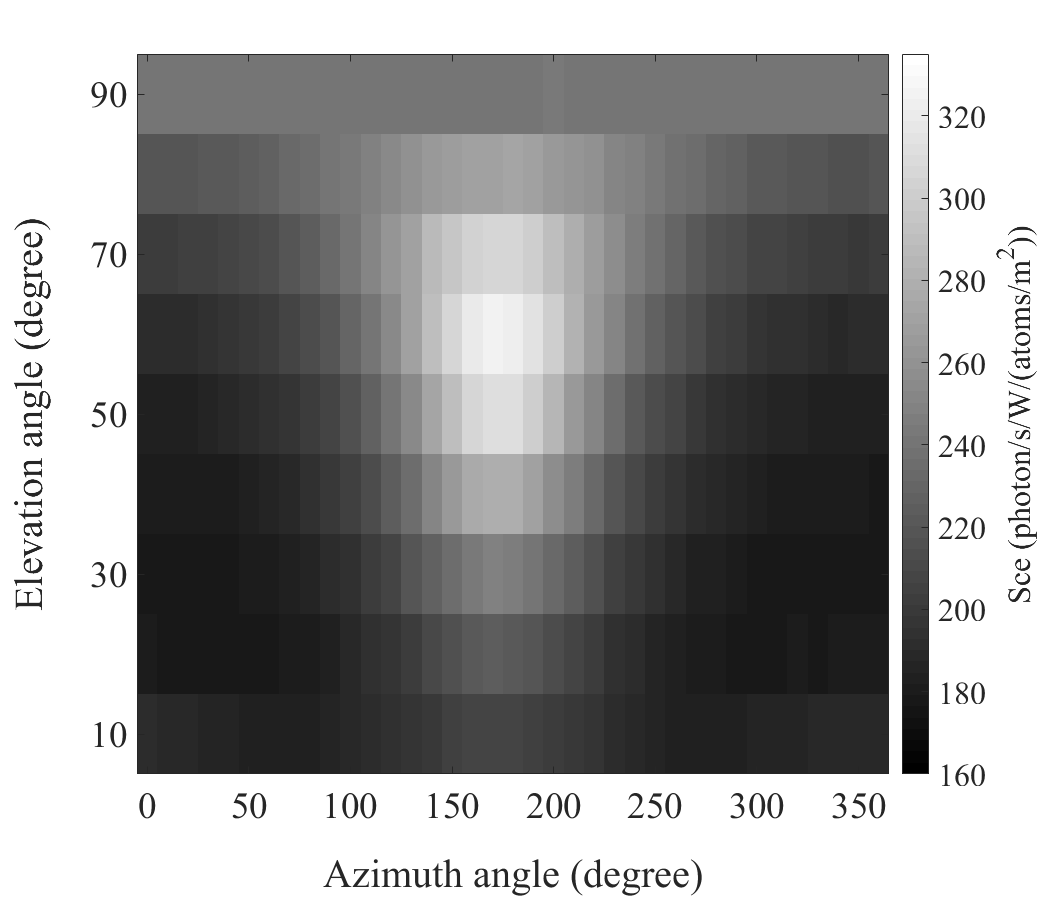}
\end{minipage}
\begin{minipage}[t]{0.5\linewidth}
\centering
\text{Lenghu}
\includegraphics[width=\linewidth]{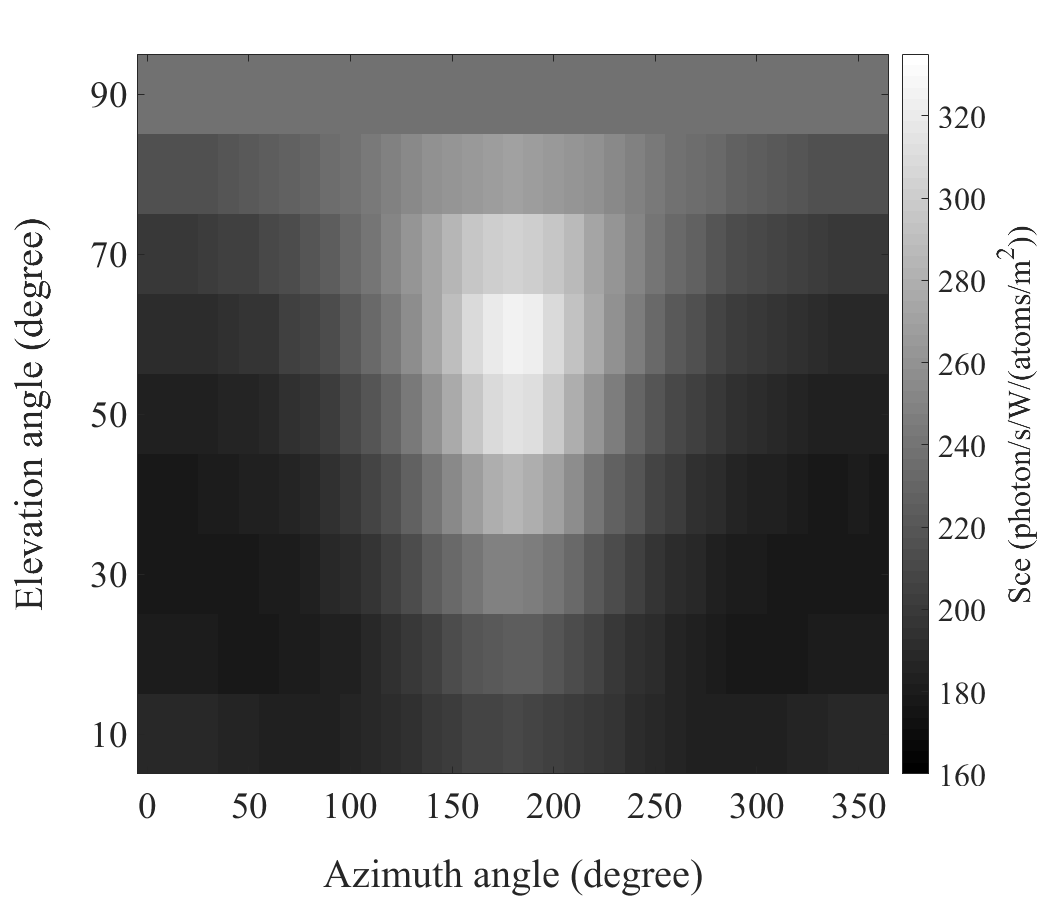}
\end{minipage} 

\begin{minipage}[t]{0.5\linewidth}
\centering 
\text{Muztagh Ata}
\includegraphics[width=\linewidth]{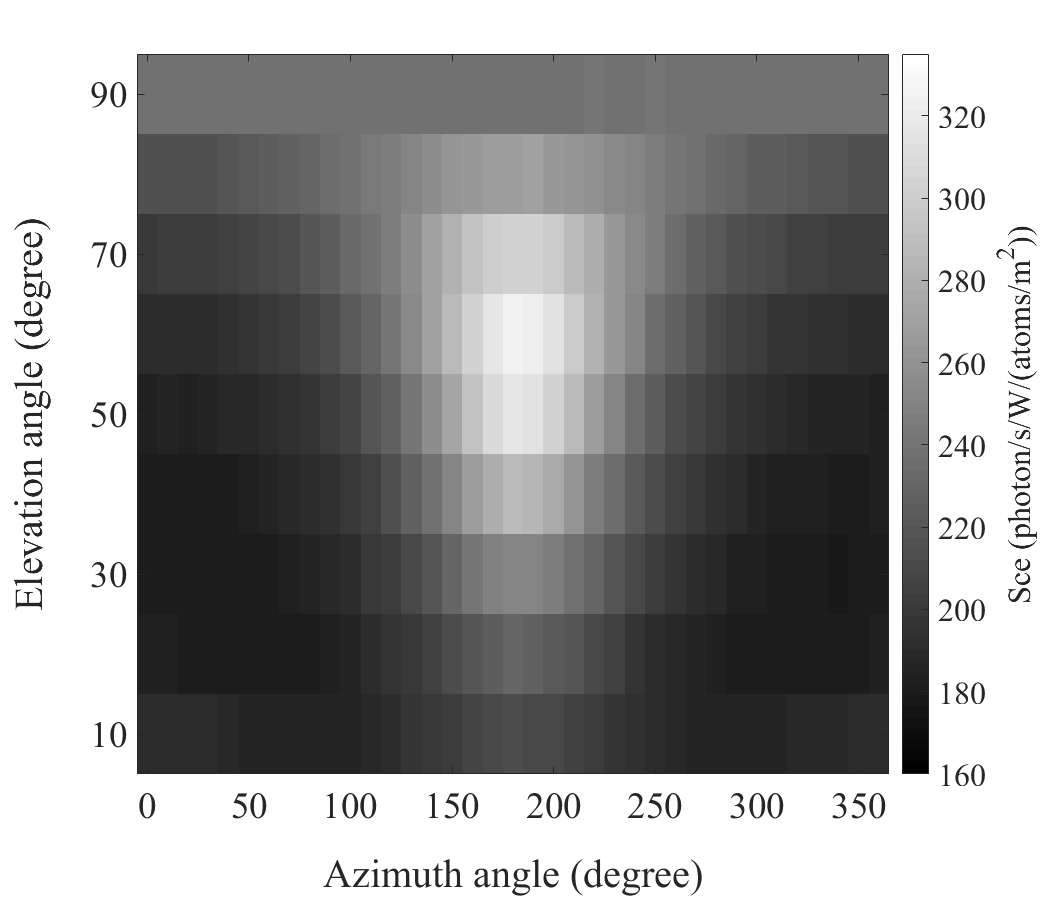}
\end{minipage}
\begin{minipage}[t]{0.5\linewidth}
\centering 
\text{Ali}
\includegraphics[width=\linewidth]{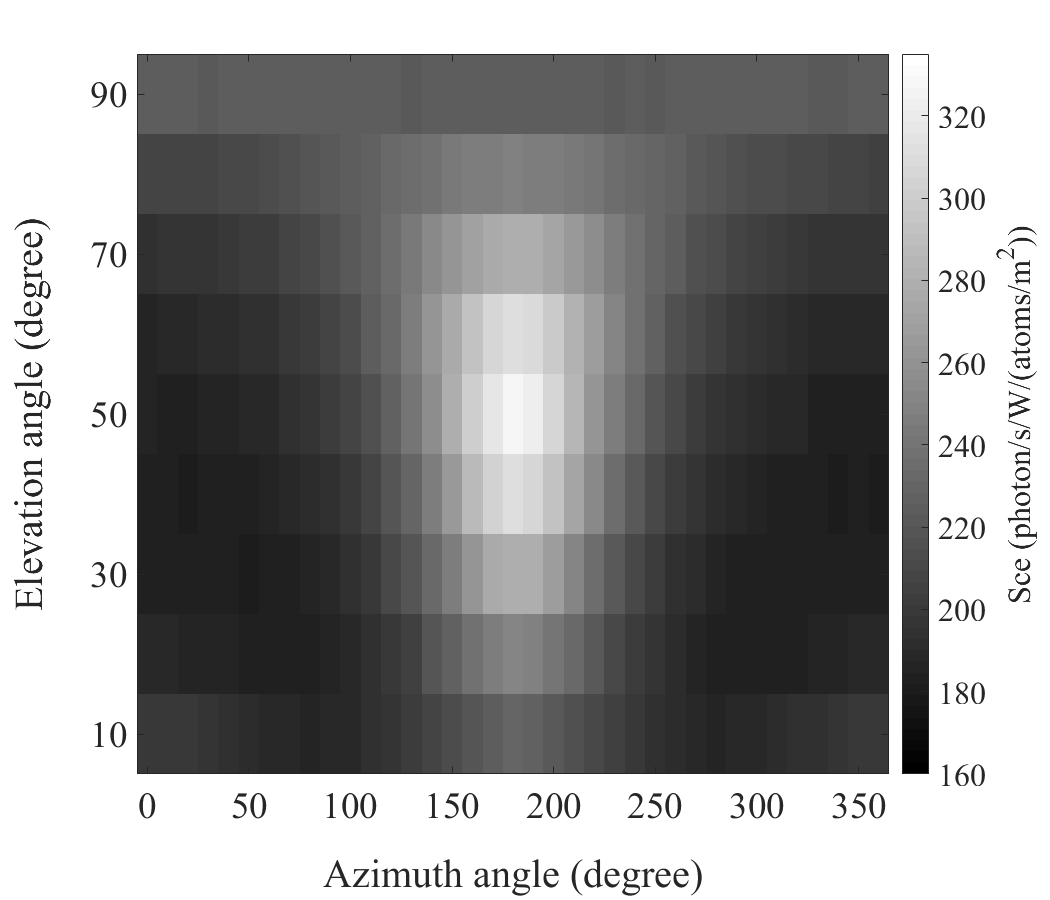}
\end{minipage} 

\text{ \hspace{73pt} Gaomeigu}
\\
\begin{minipage}[t]{\linewidth}
\includegraphics[width=0.5\linewidth]{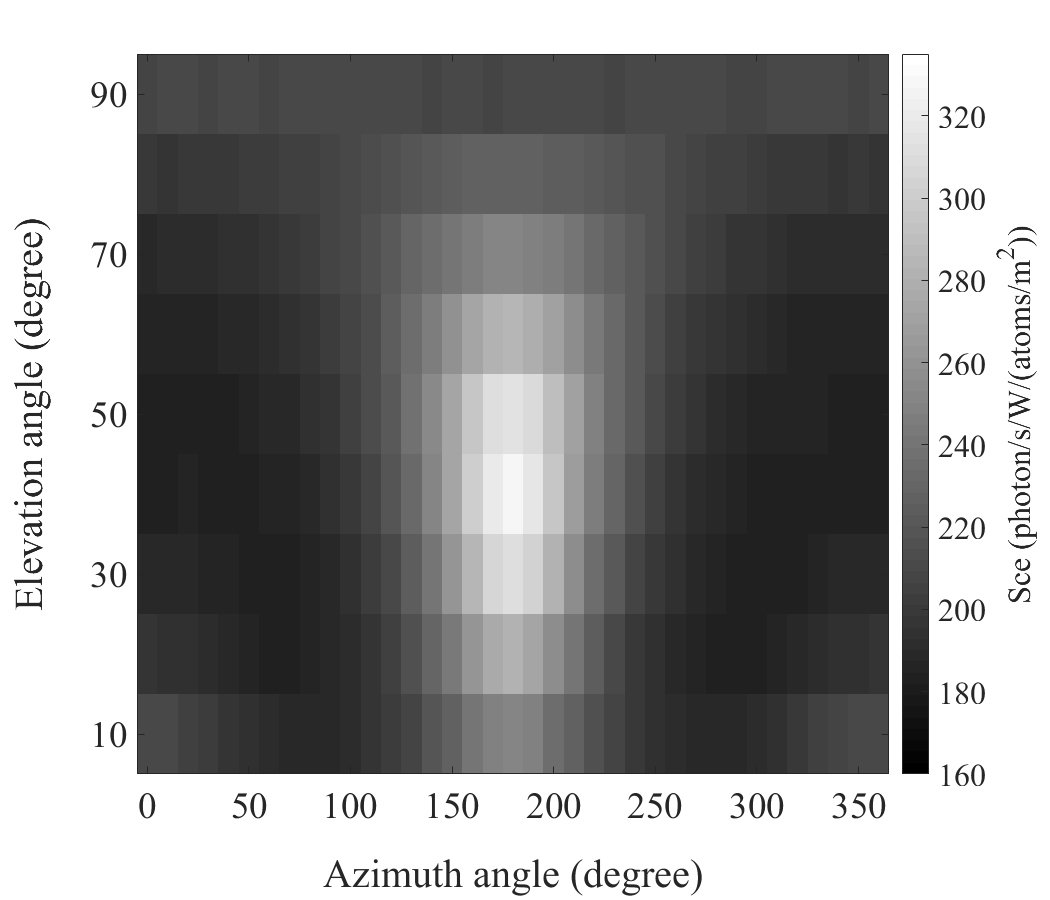}
\end{minipage}
\caption{Coupling efficiency distribution across the sky}
\label{figure: coupling efficiency change with laser direction}
\end{figure}

As mentioned in the previous section, during an astronomical observation, the direction of laser guide star would change with the pointing direction of a telescope. Thus, the angle between the laser direction and the local geomagnetic field vector will change, and this will cause continuous change of coupling efficiency because of the varying influence of the geomagnetic field. In addition, for slanted pointing directions, the increasing airmass will reduce both the projected laser power and the return photon flux. Therefore, the coupling efficiency has a peculiar spatial distribution on sky with different pointing directions. In this case, we did our simulation with a constant 20 W laser output power. For every laser pointing direction, we calculated the angle between the laser and the geomagnetic field vector, and applied this value into our simulation along with projected power density and size of spot due to airmass and slanted projection. The atmospheric transparency for every site is assumed to be 90\%. The calculated on-sky coupling efficiency results are shown in Figure \ref{figure: coupling efficiency change with laser direction}.

In each picture of Figure \ref{figure: coupling efficiency change with laser direction}, the X-axis represents the change of the azimuth angle, with the true north at 0\degree and the east at  90\degree. The Y-axis represents the change of the elevation angle, with the horizontal direction at 0\degree and the zenith at 90\degree. The variation of the coupling efficiency with laser direction at can be summarized as follows:

\begin{itemize}
    \item The coupling efficiency is almost symmetric along 180\degree of azimuth angle. When the azimuth angle changes from north to south (0\degree-180\degree), the coupling efficiency increases gradually and then decreases when the azimuth angle changes from south to north (180\degree-360\degree);
    \item The coupling efficiency increases first and then decreases with the increase of elevation angle , but the elevation angles of the maximum coupling efficiency are different among the sites;
    \item In Xinglong, Lenghu and Muztagh Ata, the coupling efficiency reaches maximums of 325.3, 327.2 and 325.5 (photons/s/W/(atoms/$m^{2}$)) at elevation angle of 60\degree and azimuth angle of about 180\degree; In Ali, the coupling efficiency reaches a maximum of 326.1 (photons/s/W/(atoms/$m^{2}$)) at elevation angle of 50\degree and azimuth angle of about 180\degree. In Gaomeigu, the coupling efficiency reaches a maximum of 328.7 (photons/s/W/(atoms/$m^{2}$)) at elevation angle of 40\degree and azimuth angle of about 180\degree. Statistics of the results are summarized in Table \ref{table: nummerical information at five sites}. The mean coupling efficiency is the average in all direction, and the sodium column density is adopted the average of 12 months when we calculate the corresponding V magnitude of mean coupling efficiency.
\end{itemize}

%\begin{table}[]
%\centering
%\begin{tabular}{|c|c|c|c|c|c|c|c}
%\hline
%Site  & \makecell[c]{The elevation \\ angle of the \\ maximum \\ coupling \\ efficiency} & 
%\makecell[c]{Mean \\ coupling \\ efficiency * \\ (photons/s/W/ \\ (atoms/$m^2$))}  &
%\makecell[c]{V magnitude \\ of the mean \\ coupling \\ efficiency}    &
%\makecell[c]{Maximum \\ coupling \\ efficiency \\ (photons/s/W/ \\ (atoms/$m^2$))}    &   
%\makecell[c]{V magnitude \\ of the \\ maximum \\ coupling \\ efficiency}  &
%\makecell[c]{Minimum \\ coupling \\ efficiency \\ (photons/s/W/ \\ (atoms/$m^2$))}    &   
%\makecell[c]{V magnitude \\ of the \\ minimum \\ coupling \\ efficiency}  \\ \hline
%Xinglong & 70\degree  & 135.3  &  7.424   &  257.5  &   6.724   &  17.31  &  9.656\\ \hline
%Lenghu & 70\degree   & 134.5 &  7.430   &  255.5   &   6.733  &  17.33 & 9.654 \\ \hline
%Muztagh Ata & 70\degree   &  135.1   &  7.425   &  254.7  &   6.737  &  17.50  &  9.644  \\ \hline
%Ali & 60\degree    & 131.3   &  7.456   &  238.4   &   6.808   &   17.69   &  9.632  \\ \hline
%Gaomeigu & 60\degree  & 126.6 &  7.496   &  217.7  &   6.907   &   17.82   &  9.624   \\ \hline
%\end{tabular}
%\caption{Numerical information about coupling efficiency (* mean coupling efficiency of all direction) %\textcolor{red}{[at five sites (Xinglong, Lenghu, Muztagh Ata, Ali, Gaomeigu)]}}
%\label{table: nummerical information at five sites}
%\end{table}

\begin{table}[]
\centering
\begin{tabular}{|c|c|c|c|c|c|c|}
\hline
Site  & \makecell[c]{The elevation \\ angle of the \\ maximum \\ coupling \\ efficiency} & 
\makecell[c]{Mean \\ coupling \\ efficiency * \\ (photons/s/W/ \\ (atoms/$m^2$))}  &
\makecell[c]{V magnitude \\ of the mean \\ coupling \\ efficiency}    &
\makecell[c]{Maximum \\ coupling \\ efficiency \\ (photons/s/W/ \\ (atoms/$m^2$))}    &   
\makecell[c]{Minimum \\ coupling \\ efficiency \\ (photons/s/W/ \\ (atoms/$m^2$))}   & 
\makecell[c]{Elevation and \\ azimuth angle \\ of minimum \\ coupling \\ efficiency}\\ \hline 
Xinglong & 60\degree  & 218.3  &  6.904   &  325.3    &  177.2  &  (El:30\degree Az:330\degree) \\ \hline
Lenghu & 60\degree   & 217.5 &  6.908  &  327.2   &  177.4  &  (El:30\degree Az:40\degree) \\ \hline
Muztagh Ata & 60\degree   &  218.9   & 6.901   &  325.5  &  178.7  &  (El:30\degree Az:340\degree) \\ \hline
Ali & 50\degree    & 216.1   &  6.915  &  326.1   &   181.4  &  (El:30\degree Az:50\degree)  \\ \hline
Gaomeigu & 40\degree  & 212.1 &  6.935   &  328.7   &   182.5  &  (El:30\degree Az:50\degree)  \\ \hline 
\end{tabular}
\caption{Numerical information about coupling efficiency (* mean coupling efficiency in all direction)}
\label{table: nummerical information at five sites}
\end{table}

\section{Conclusion}
\label{section: conclusion}

In this paper, we have studied the performance of sodium laser guide stars generated by a 20W class quasi-CW pulsed laser at five typical sites of China. From Table \ref{table: nummerical information at five sites}, the overall annual mean performances of sodium guide star at these sites are very similar, and the similarity can be summarized as due to that the brightness of the guide star is strongly relying on the seasonal variation of sodium column density and laser pointing direction rather than geographical locations. However, for AO design concerns, coupling efficiency's variation in different months with different pointing angles are more important because it determines the performance that the AO system could function in specific night of year at a particular angle. The seasonal photon flux distribution of LGS shows higher value in winter and lower in summer. The photon return in summer is only about 44\% (at Xinglong site) - 53\% (at Gaomeigu site) of that in winter (see Figure \ref{figure: photon flux and V magnitude}), which renders a difference of 0.882-0.685 in V magnitude. For different laser pointing direction, the coupling efficiency changes most significantly with the azimuth angle at elevation angle from 40\degree to 60\degree. In addition, the variation of coupling efficiency with elevation angle is also prominent, which shows that the elevation angle for the maximum coupling efficiency is higher with increasing latitude as shown in Figure \ref{figure: coupling efficiency change with laser direction} and Table \ref{table: nummerical information at five sites}. With both effects combined, the variation of sodium LGS's brightness is not negligible in AO system's design process. We hope that this work could shed some light for future applications of sodium laser guide star.

\section{Acknowledgement}
\label{section: ackowledgement}
The research is supported by the Operation, Maintenance and Upgrading Fund for Astronomical Telescopes  and  Facility  Instruments,  budgeted  from  the  Ministry  of Finance of China (MOF) and administrated by the Chinese Academy of Sciences (CAS).

\label{lastpage}


\begin{thebibliography}{99}
%% you can type \apj for ApJ, \aap for A&A, \apss for Ap&SS, etc. Please consult
%% the macro chjaa.cls. You can also find them in aasguide.tex (AASTeX for ApJ, AJ, PASP)
%% Please follow the format of ChJAA's reference list

  \bibitem[Bian et al.(2020)]{Bian+etal+2020} Bian, Q., Bo, Y., Zuo, J. W., et al., 2020, OptL, 45, 7, 1818

  \bibitem[Bian et al.(2016a)]{Bian+etal+2016a} Bian, Q., Bo, Y., Zuo, J. W., et al., 2016, OptL, 41, 8, 1732

  \bibitem[Bian et al.(2016b)]{Bian+etal+2016b} Bian, Q., Zuo, J. W., Guo, C., et al., 2016, LaPhy, 26, 095005

  \bibitem[Bian et al.(2017)]{Bian+etal+2017} Bian, Q., Zong, Q. S., Chang, J. Q., et al., 2017, IPTL, 29, 2095

  \bibitem[Calia et al.(2014)]{Calia+etal+2014} Calia, D. B. , Hackenberg, W. , Holzlöhner, R., et al., 2014, AdOT, 3, 3, 345
  
  \bibitem[Chin et al.(2016)]{Chin+etal+2016} Chin, J. C. Y., Wizinowich, P., Wetherell, E., et al. 2016, Proc.SPIE, 9909, 99090S
  
  \bibitem[Feng et al.(2016)]{Feng+etal+2016} Feng, L., Shen, Z. X., Xue, S. J., et al., 2016, RAA(Research in Astronomy and Astrophysics), 16, 144

  \bibitem[Feng et al.(2015)]{Feng+etal+2015} Feng, L., Kibblewhite, E., Jin, K., et al., 2015, Proc. SPIE, 9678, 96781B

  \bibitem[Feng et al.(2020)]{Feng+etal+2020} Feng, L., Hao, J. X., Cao, Z. H., et al., 2020, RAA(Research in Astronomy and Astrophysics), 20, 80

  \bibitem[Feng \& Hao(2020)]{Feng+Hao+2020} Feng, L., Hao, J. X. 2020, RAA(Research in Astronomy and Astrophysics), 20, 79

  \bibitem[Fussen et al.(2004)]{Fussen+etal+2004} Fussen, D., Vanhellemont, F., Bingen, C., et al., 2004, GeoRL, 31, L24110
  
  \bibitem[Fussen et al.(2010)]{Fussen+etal+2010} Fussen, D., Vanhellemont, F., Tétard, C., et al., 2010, ACP, 10, 9225
  
  \bibitem[Hardy(1998)]{Hardy+1998} Hardy, J. W., 1998, Adaptive Optics for Astronomical Telescopes (New York: Oxford University Press)
  
  \bibitem[Holzlöhner et al.(2010)]{Rochester+etal+2010} Holzlöhner, R., Rochester, S. M., Calia, D. B., et al., 2010, A\&A, 510
  
  \bibitem[Jin et al.(2015)]{Jin+etal+2015} Jin, K., Wei, K., Feng, L., et al., 2015, PASP, 127, 749
  
  \bibitem[Jin et al.(2014)]{Jin+etal+2014} Jin, K., Wei, K., Xie, S., et al., 2014, Proc. SPIE, 9148, 91483L
  
  \bibitem[Kibblewhite(2000)]{Kibblewhite+2000} Kibblewhite, E. J., 2000, in Laser Guide Star  Adaptive  Optics for  Astronomy,  Ageorges, N. \&  Dainty, C., (Kluwer Academic Publishers), 51
  
  \bibitem[Kibblewhite(2008)]{Kibblewhite+2008} Kibblewhite, E. J., 2008, Proc. SPIE 7015, 70150M
  
  \bibitem[Langowski et al.(2017)]{Langowski+etal+2017} Langowski, M. P., Savigny, C., Burrows, J. P., et al., 2017, AMT, 10, 2989
  
  \bibitem[Li et al.(2021)]{Li+etal+2021} Li, H. Y., Feng, L., Wang, J. L., et al., 2021, PASP, 133, 1021
  
  \bibitem[Liu et al.(2012)]{Liu+etal+2012} Liu, L. Y., Yao, Y. Q., Vernin, J., et al., 2012, in Proc. of SPIE, 8444, Ground-based and Airborne Telescope IV, 844464
  
  \bibitem[Liu et al.(2015)]{Liu+etal+2015} Liu, L. Y., Yao, Y. Q., Vernin, J., et al., 2015, Journal of Physics: Conference Series, 595, 012019
  
  \bibitem[Liu et al.(2016)]{Liu+etal+2016} Liu, Y., Song, T. F., Zhang, X. F., et al., 2016, in Proc. of IAU, 320, Solar and Stellar Flares and their Effects on Planets, 447
  
  \bibitem[Moussaoui et al.(2008)]{Moussaoui+etal+2008} Moussaoui, N., Holzlöhner, R., Hackenberg, W., et al., 2008, Proc. SPIE, 7015, 70152W
  
  \bibitem[Moussaoui et al.(2009)]{Moussaoui+etal+2009} Moussaoui, N., Holzlöhner, R., Hackenberg, W., et al., 2009, A\&A, 501, 793
  
  \bibitem[Otarola et al.(2016)]{Otarola+etal+2016} Otarola, A., Hickson, P., Gagne, R., et al., 2016, JAI, 5, 1
  
  \bibitem[Qian et al.(2015)]{Qian+etal+2015} Qian, X., Yao, Y. Q., Wang, H. S., et al., 2015, PKAS, 30, 695
  
  \bibitem[Wang et al.(2015)]{Wang+etal+2015} Wang, H. S., Yao, Y. Q., Liu, L. Y., et al., 2015, Journal of Physics: Conference Series, 595, 012037
  
  \bibitem[Wei et al.(2012)]{Wei+etal+2012} Wei, K., Bo, Y., Xue, X.H., et al. 2012, Proc.SPIE, 8447, 84471R
  
  \bibitem[Wu et al.(2016)]{Wu+etal+2016} Wu, N., Liu, Y., \& Zhang, H. M. 2016, AcASn, 57, 729
  
  \bibitem[Xu et al.(2020)]{Xu+etal+2020} Xu, J., Esamdin, A., Hao, J.X., et al., 2020, RAA(Research in Astronomy and Astrophysics), 20, 86
  
  \bibitem[Yao et al.(2013)]{Yao+etal+2013} Yao, Y. Q, Wang, Y. P., Liu, L. Y., et al., 2013, in NARIT Conf. Ser., 1, the 11th Asian-Pacific Regional IAU Meeting 2011, ed. Komonjinda, S. (IAU), 1,  http://conference.narit.or.th/ncs/APRIM2011\_proceedings/S7/S7P05.pdf
  

\end{thebibliography}
\end{document}